\documentclass[10pt,twocolumn,superscriptaddress,showpacs,floatfix]{revtex4-2}
\usepackage{graphics}
\usepackage{graphicx}
\usepackage{bm}
\usepackage{amsmath}
\usepackage{amsfonts}
\usepackage{amssymb}
\usepackage{latexsym}
\usepackage[dvipsnames]{xcolor}
\usepackage{hyperref}
\usepackage{float}
\begin{document}

\title{Quantum mechanics in configuration space in context}

\author{Arwa Bukhari} 
\affiliation{School of Physics and Astronomy, University of Leeds, Leeds LS2 9JT, United Kingdom.}
\affiliation{School of Physics, Umm Al-Qura University, Makkah, Saudi Arabia.}
\author{Margherita Moro}
\affiliation{School of Philosophy, Religion and History of Science, University of Leeds, Leeds LS2 9JT, United Kingdom.}
\author{Max J.~Davies} 
\affiliation{School of Physics and Astronomy, University of Leeds, Leeds LS2 9JT, United Kingdom.}
\author{Alastair Wilson}
\affiliation{School of Philosophy, Religion and History of Science, University of Leeds, Leeds LS2 9JT, United Kingdom.}
\author{Almut Beige}
\affiliation{School of Physics and Astronomy, University of Leeds, Leeds LS2 9JT, United Kingdom.}

\date{\today}

\begin{abstract}
The formalism of {\em quantum mechanics in configuration space} (QMCS) [Bukhari {\em et al.}, New J. Phys. {\bf 27}, 084501 (2025)] has been introduced to unify the modelling of photons and quantum mechanical particles. In this paper, we illustrate that QMCS also increases the continuity between quantum and classical mechanics, since it can be used to obtain both their predictions. Being based on a physically motivated quantisation of Newtonian mechanics, QMCS makes a clear distinction between the velocity observable $\hat v$ and the generator for spatial translations, i.e. the momentum operator $\hat p$, of mechanical particles. In contrast, standard quantum mechanics, which is based on Hamiltonian mechanics, relates velocity directly to momentum rather than treating them as independent observables. Despite having many formal similarities with Hilbert space formulations of  classical mechanics, QMCS is a quantum theory. To illustrate this and to further explore the relations between QMCS and alternative approaches, this paper studies quantum mechanical particles moving in free space as an example.
\end{abstract}

\maketitle

\section{Introduction} \label{sec:1}

Physics often offers many different descriptions of the same phenomenon. The question arises of how models of the same phenomenon belonging to different theories are related, and which criteria of inter-theory consistency to adopt (for a variety of options, see \cite{Palacios}). For example, classical mechanics (CM) and quantum mechanics (QM) are two different frameworks \cite{Wallace2020} describing the behaviour of mechanical particles. These frameworks are generally understood as complementing one another, rather than as clashing: they characterise different aspects of particle behaviour, and so they can be used alongside one another. Furthermore, they can be ordered with respect to ontological fundamentality: we typically say that the more fundamental aspects of the behaviour of mechanical particles are described by the more fundamental theory (QM), with the less fundamental theory (CM) describing \textit{derivative} or \textit{emergent} aspects. This ontological notion of fundamentality may be contrasted with a conceptual notion of fundamentality, understood in terms of closeness to our ordinary, common-sense, conceptual framework.

\begin{figure}[t]
\centering
\includegraphics[width=0.5 \linewidth] {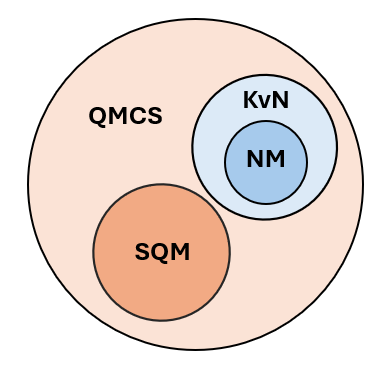}
\caption{Schematic view of the possible states of mechanical particles moving in free space according to quantum mechanics in configuration space (QMCS), to standard quantum mechanics (SQM) and to Hilbert space formalism of classical mechanics, known as Koopman-von Neumann mechanics (KvN) \cite{Sudarshan,Mauro}\textemdash which includes states of Newtonian mechanics (NM). The blue and the orange colours refer to classical and to quantum theories, respectively. The Hilbert space of QMCS has been constructed such that it contains states with the same dynamics as the states $|\psi \rangle$ and $(x,v)$ of SQM and of NM. Moreover, considering the density matrices $\rho$ in Eq.~(\ref{E1.3}), QMCS can reproduce the predictions of KvN. In addition, QMCS supports additional quantum states which are not covered by the above mentioned alternative theories (cf.~e.g.~Eq.~(\ref{E1.4})).} \label{fig:0}
\end{figure}

How is it possible for distinct theories to successfully describe the same physical system? The most familiar answer draws on \textit{inter-theoretic reduction}. In general, where T (the \textit{reduced theory}) is a theory superseded by T' (the \textit{reducing theory}), there are some constraints on how T and T' relate. Most importantly, both the successes and the limitations of the reduced theory need to be explained by the reducing theory. We may need to use various approximations, idealizations, and other modelling assumptions to obtain this explanation. However, once established, this explanatory relation may justify the continued use of the reduced theory within a certain domain, even where the superseding theory is also available \cite{Fletcher, RobertsonWilson}.

A key role which may be played by the reduced theory is that of \textit{epistemic guide} to the reducing theory \cite{Nickles, Feintzeig}. That is, the reduced theory can work as a heuristic tool by informing our choice among the range of variables and mathematical relations in candidate successor theories. As already mentioned, any adequate successor must be capable of yielding the reduced theory through a suitable modelling procedure, and hence it must be capable of explaining that theory's effectiveness (and its limits). On top of that, the superseded theory might present some particularly appealing features that one tries to preserve when exploring possible reducing theories. For instance, some structural feature of the models or the ontology of the reduced theory might continue to be regarded as both theoretically and pragmatically fruitful and intuitive when confronted with questions about the underlying physical reality and the role of measurement. It is worth noting that this does not mean that the superseded theory strictly constrains \textit{all} aspects of the reducing theory: the more fundamental theory aims to describe new aspects of the same phenomenon, and might therefore require a completely new methodology in order to do so. Nevertheless, our choice of which reduced theory to focus on can help determine which reducing theory we end up adopting. This epistemic role is particularly illuminating when exploring how the classical and quantum descriptions of mechanical particles relate.

QM aims to be a reducing theory with respect to CM. Otherwise, it would be possible to contradict QM by simply performing a classical experiment. In an initial attempt to identify the quantum-classical transition, Ehrenfest  \cite{Ehrenfest} pointed out that expectation values of quantum mechanical observables evolve in the same way as certain classical quantities in CM. However, closer examinations of Ehrenfest's theorem revealed that it only applies in a limited number of situations \cite{Ballentine,Wheeler}. For example, the theorem holds for a particle moving in free space and for harmonic oscillators. In general, its validity is only guaranteed in the presence of highly localised wave functions which remain localised at all times. Unfortunately, as shown in Refs.~\cite{Heger,Heger2}, this is usually not the case in standard quantum mechanics (SQM) \cite{Griffiths,Sakurai,Liboff}. Thus, Ehrenfest's theorem remains severely limited with respect to the task of bridging the gap between SQM and CM.

Textbooks on SQM typically adopt a canonical quantisation approach, which imposes a non-zero commutator relation between canonical variables \cite{Klauder}. Nevertheless, several alternative quantisation schemes have been put forward, including geometric quantisation \cite{Carosso} and integral quantisation \cite{Ali}. In geometric quantisation, classical observables existing on a symplectic manifold are mapped onto operators which act on a Hilbert space, through a pre-quantisation procedure. The resulting Hilbert space is considered too large and is subsequently reduced by projecting out unwanted degrees of freedom, thereby regaining the desired canonical commutation relations. Similarly, integral quantisation constructs quantum observables via integral transforms over the parameter space of CM, offering greater flexibility, yet at the cost of conceptually ambiguous mappings. However, none of the above approaches significantly improves the continuity between quantum and classical descriptions. A more widely-accepted way of linking the predictions of quantum and classical formalisms is to invoke decoherence, which alters the dynamics of quantum systems such that they become approximately classical in certain respects. However, the mechanisms that cause decoherence to occur often remain unclear, and its interpretive significance is disputed \cite{Zurek,Brody,Brukner}. 

Recently, Bukhari {\em et al.}~\cite{Bukhari} introduced an alternative formulation of quantum mechanics, namely {\em quantum mechanics in configuration space} (QMCS), in an attempt to enhance wave-particle duality, i.e.~to unify the theoretical modelling of photons and of mechanical particles. As we demonstrate in this paper, QMCS likewise increases the continuity between QM and CM (cf.~Fig.~\ref{fig:0}).
For example, QMCS contains quantum states with well-defined positions and velocities which can be used to represent  mechanical particles. Others represent the quantum superposition states $|\psi \rangle$ of SQM. In addition, there are additional states that are covered neither by CM nor by SQM. Whether these states can be physically realised needs to be decided by experiments \cite{Davies}. Our hope is that the enlarged Hilbert space of QMCS will allow us to gain new insights into physical systems which evade simple quantum descriptions, such as gravity, relativity, chemistry and chaos.

One reason why identifying the quantum-classical transition is not straightforward is that CM itself admits several formulations, with the three most common being Newtonian, Lagrangian and Hamiltonian mechanics \cite{Goldstein}. Each formulation highlights different aspects of the behaviour of classical mechanical particles and is better suited for the description of certain physical situations than others \cite{gregory,Dirac}. As this paper emphasises, some formulations are more physically {\em perspicuous} than others, in that their characteristic quantities more closely correspond to what we take to be fundamental (in the ontological sense). Two theories may both be used to correctly describe reality, but the more perspicuous of the theories will more directly represent the physical reality through using characteristic quantities which correspond more directly to the ontological elements. Perspicuity should be contrasted with mathematical elegance: while physical perspicuity tracks adherence to the underlying physical ontology, mathematical elegance tracks aesthetic and epistemic values related to the simplicity of the formalism. While SQM is based on Hamiltonian mechanics, the starting point for the derivation of the QMCS formalism is Newtonian mechanics which, as we argue in this paper, is the most physically perspicuous formulation of CM.

Furthermore, while SQM is based on a canonical quantisation procedure, QMCS employs an alternative approach: {\em physically motivated quantisation} \cite{Bennett,Jake,Daniel}. A physically motivated quantisation procedure is one which starts from a description of a physical domain that is taken to be perspicuous, and transforms that representation in a way which preserves the representational role of the key quantities associated with that perspicuous description. In the present case, we take our approach to quantisation to be physically more perspicuous than canonical quantisation, since its starting point is the configuration space of the corresponding classical system (as opposed to the smaller phase space). More specifically, Newtonian mechanics describes mechanical particles moving in one dimension by specifying their position $x$ and their velocity $v$. Since the corresponding $(x,v)$ states of CM are {\em distinguishable}, QMCS promotes them to pairwise orthogonal $|x,v\rangle$ quantum states. These form an orthonormal basis in the Hilbert space $\mathcal{H}$ of the formalism, i.e.~in the quantum {\em configuration space}, which contains all possible states of quantum mechanical particles.\footnote{In physics, there are different definitions of the term {\em configuration space}. For example, in CM, this term often only refers to the position space, since the velocity $v$ of a mechanical particle coincides with the time derivative of $x$. Here we take the point of view that the configuration space should contain all independent variables that are needed to describe the state of a physical system at a given time $t$, i.e.~the position $x$ as well as the velocity $v$. Notice also that the term {\em phase space} refers to the position-momentum space of mechanical particles which is in general different from the position-velocity space that we consider here.} 
The state vector $|\psi \rangle$ of quantum mechanical particles in a pure state is therefore of the general form
\begin{eqnarray} \label{E1.2}
|\psi \rangle = \iint_{{\mathbb{R}}^2} {\rm d}x \, {\rm d}v \, \psi(x,v,t) \, |x,v \rangle 
\end{eqnarray}
with $\psi(x,v,t)$ denoting the corresponding wave function.
Here, the integrals are taken over the parameter set ${\mathbb{R}}^2 = \{(x,v)|x,v \in \mathbb{R}\}$ with $x $ and $v$ representing position and velocity, respectively.
By applying the rule of SQM for the construction of observables, one can then construct position and velocity observables, $\hat x$ and $\hat v$, respectively, and calculate their expectation values in a given quantum state $|\psi\rangle$. In addition, we identify generators for changes in position, in velocity, and in time, i.e.~the momentum operator $\hat p$, the acceleratum $\hat a$ and the dynamical Hamiltonian $\hat H_{\rm dyn}$ \cite{Bukhari}. In this paper, we remain as neutral as possible on questions of interpretation of quantum theory, and in particular on the proper interpretation of quantum probabilities. We intend QMCS to be compatible with any interpretation which does not introduce new ontology or new physics for measurement interactions\textemdash for instance, with Everettian interpretations \cite{Wallace2012}, relational interpretations \cite{Rovelli1996}, and QBism \cite{FuchsSchack2011}.

QMCS bears some resemblance to Bohmian mechanics \cite{Bohm, Durr1}; as the two approaches share certain features. For instance in associating well-defined positions and velocities with particle states at all times. However, the way this is achieved is very different in QMCS and in Bohmian mechanics, and the two theories should not be conflated. QMCS does not add additional ontology beyond what is represented by the quantum state of the system, and position and velocity remain ordinary physical quantities. Bohmian mechanics adds additional point particles to the ontology of the quantum state. It also treats position and velocity as different in kind from other physical quantities, as attaching to the particles directly rather than to the quantum state. Moreover, Bohmian mechanics lacks a consistent extension to quantum field theory without the introduction of further mathematical structure \cite{Wallace2020}.

At a first glance, QMCS shares many similarities with Hilbert space formulations of classical mechanics. The historical development of these approaches, which are closely related to the method of classical wave functions, has been traced by \cite{Barandes}. The most prominent example is the mechanics of Koopman and von Neumann (KvN) \cite{Koopman,vonNeumann}. In KvN mechanics, classical particles are described by state vectors $|\psi\rangle$ of the same form as those in Eq.~(\ref{E1.2}). Furthermore, the dynamics of these state vectors are governed by a Schrödinger-like equation, with generators that play roles analogous to those in QMCS, namely $\hat p$, $\hat a$ and $\hat H_{\rm dyn}$. However, KvN imposes a superselection rule to ensure
classicality  \cite{Sudarshan,Mauro}. Hence the {\em interpretation} of the state vectors $|\psi \rangle$ in Eq.~(\ref{E1.2}) is in fact different in KvN and QMCS. In KvN, state vectors represent statistical ensembles of classical mechanical particles. In the language of modern quantum physics, such states are usually represented by density matrices $\rho$ of the form\footnote{The original formulation of KvN replaces the 
velocity $v$ by the momentum $p$ with $p=mv$, to obtain equations which look like the equations of motion in Hamiltonian mechanics. Moreover, KvN introduces a generator for spatial translations, which is independent of the velocity operator. A more detailed discussion of KvN can be found in Section \ref{sec4}.}
\begin{eqnarray} \label{E1.3}
\rho = \iint_{{\mathbb{R}}^2} {\rm d}x \, {\rm d}v \, |\psi(x,v,t)|^2 \, |x,v \rangle \langle x,v| \, .
\end{eqnarray}
Using a different interpretation, the states $|\psi \rangle$ of 
QMCS represent genuine quantum superposition states. In addition, QMCS supports statistical mixtures of coherent superpositions with density matrices $\rho$ of the general form
\begin{eqnarray} \label{E1.4}
\rho = \iint_{{\mathbb{R}}^2} {\rm d}x \, {\rm d}v \iint_{{\mathbb{R}}^2} {\rm d}x' \, {\rm d}v' \, \rho(x,v,x',v',t) \, |x,v \rangle \langle x',v'| \,. \notag \\
\end{eqnarray}
The $\rho(x,v,x',v',t)$  
in this equation are (complex) matrix elements with
$\rho(x,v,x',v',t) = \rho(x',v',x,v,t)^*$.
The above density matrix includes Eq.~(\ref{E1.3}) as a special case. If the proponents of KvN had relaxed its superselection rule and had not restricted the approach to apply only to the description of classical statistical ensembles \cite{Mauro}, then the result would have been a {\em quantum} theory similar to the QMCS quantum formalism.

This paper is organised as follows. Section~\ref{sec:2} reviews the three main formulations of CM, namely Newtonian, Langrangian and Hamiltonian mechanics, and highlights their distinct features. Section~\ref{sec:3} reviews the basic formalism of QMCS and introduces its basic observables and generators. Section \ref{sec4} studies the dynamics of a particle moving in free space as an example. In addition, we discuss the relations of QMCS to Newtonian mechanics, to SQM and to KvN for this specific case. Finally, we review our findings in Section~\ref{sec5}. 

\section{The main formulations of classical mechanics} \label{sec:2}

\begin{figure}[t]
\centering
\includegraphics[width=0.7 \linewidth] {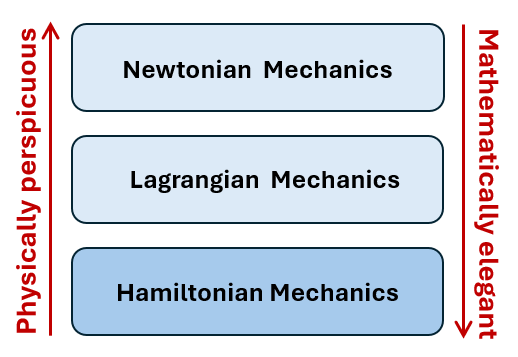}
\caption{An illustration of the three main formulations of classical mechanics and how physically perspicuous we take them to be. As we argue below, Newtonian mechanics provides the most perspicuous description of classical mechanical particles. Nonetheless, one can derive the mathematically more elegant framework of Lagrangian mechanics from it, but at the cost of physical perspicuity. Furthermore, one can adopt the (sometimes irreversible) Legendre transform to progress to the evermore mathematically elegant Hamiltonian mechanics. However, Hamiltonian mechanics is far less physically perspicuous than Newtonian mechanics (i.e., Newtonian mechanics more directly represents the properties of the underlying physical systems).} \label{fig:1}
\end{figure}

Classical mechanics (CM) is commonly formulated within three principal mathematical frameworks, each offering a distinct description of the dynamics of physical systems. These are Newtonian \cite{Newton}, Lagrangian \cite{Lagrange1997} and Hamiltonian \cite{Hamilton} mechanics which often yield the same empirical predictions and are treated as equivalent formulations of CM. However,  the three formulations differ in both their mathematical structures and their underlying conceptual commitments \cite{assumptions, HamvsLag}. The differences between Newtonian and Lagrangian mechanics on the one hand and Hamiltonian mechanics on the other are not merely formal, but reflect more meaningful distinctions regarding the fundamental structure and interpretation of the theory \cite{North}, suggesting that the claim of full equivalence requires closer examination. Some of the above formulations, we argue, provide a more perspicuous representation of the physical reality they purport to describe, than others\textemdash that is, they represent the world using a set of quantities which correspond more directly to what we take to be elements of physical reality (cf.~Fig.~\ref{fig:1}). The purpose of this section is to illustrate this claim by reviewing the main features of Newtonian, Lagrangian and Hamiltonian mechanics, thereby motivating the adoption of Newtonian mechanics as the foundation of QMCS, in contrast to SQM, which takes Hamiltonian mechanics as its starting point.

\subsection{Newtonian mechanics} 

The oldest and most familiar description of classical mechanical systems goes back to the development of calculus by Leibniz \cite{Leibniz} and Newton \cite{Newton} in the 17th century. Newtonian mechanics describes individual particles moving in one dimension in the presence of a position-dependent potential $V(x)$ by identifying their positions $x$ and their velocities $v$. The  $x$-$v$ space therefore represents the configuration space of mechanical particles. As time evolves, these particles follow unique trajectories $(x(t), v(t))$. Their dynamics are governed by Newton's second law which requires the identification of a local force $F(x)$ acting on a particle of mass $m$ such that
\begin{eqnarray} \label{E1}
F(x) = m \, \ddot x(t)\, .
\end{eqnarray}
Here $F(x) = - \partial /\partial x \, V(x)$ and the dots denote total derivatives in time. Using this notation, Eq.~(\ref{E1}) can be written as a set of two first order differential equations,
\begin{eqnarray} \label{E2}
\dot x(t) = v(t) \, , ~~~~ 
\dot v(t) = f(x) 
\end{eqnarray}
with $f(x)$ representing a local position-dependent force per mass,
\begin{eqnarray}\label{Newton5}
f(x) = - {1 \over m} \, {\partial V(x) \over \partial x} \, .
\end{eqnarray}
Eq.~(\ref{E2}) shows that $f(x)$ also equals the acceleration of the mechanical particle moving in the potential $V(x)$. 

As we shall see in Section \ref{sec:3}, QMCS stems from the guiding idea that Newtonian mechanics is the most perspicuous version of CM, and hence the most suitable starting point within CM for applying a physically motivated quantisation procedure. The primary reason for this choice is the ontological and dynamical clarity offered by Newtonian mechanics. For example, in Newtonian mechanics, mechanical particles respond to locally acting forces and behave causally \cite{causation}\textemdash with position and velocity amongst the characteristic quantities of the theory. Referring back to the methodology expounded in the Introduction, we are therefore assuming Newtonian mechanics to be \textit{conceptually} more fundamental, and hence a better epistemic guide to the description of the \textit{ontologically} more fundamental aspects of the behaviour of mechanical particles. Note that this does not preclude, however, that a quantisation procedure\textemdash not itself part of the epistemic guidelines provided by CM\textemdash can be used as a heuristic by way of getting to QMCS.

\subsection{Lagrangian mechanics} 

In his foundational treatise \cite{Lagrange1997}, Lagrange sought to eliminate geometrical constructions from CM, a goal reflected in his famous statement: “No figures will be found in this work.” Moreover, the introduction of Lagrangian mechanics was motivated by a desire to simplify the description of complex mechanical systems. To achieve these very ambitious goals, Lagrange, based his mechanics on the calculus of variations and identified a scalar function, namely the Lagrangian $\mathcal{L}(q_i,\dot q_i,t)$, which depends on the generalised coordinates $q_i$ and their time derivatives $\dot q_i$ at a given time $t$. Here, the index $i$ ranges over all the degrees of freedom of the respective mechanical system. In the case of a single particle moving along the $x$-axis, the generalised coordinates $q_i$ reduce to a single variable, namely the position $x$, and the Lagrangian $\mathcal{L}(x,\dot x,t)$ is given by 
\begin{eqnarray}\label{lagrangian}
  \mathcal{L}(x,\dot x,t) = T(\dot x) - V(x) \, .
\end{eqnarray}
Here $T(\dot x)$ and $V(x)$ denote kinetic and potential energies \cite{Goldstein}. 

As mechanical systems evolve in time, the value of the Lagrangian $\mathcal{L}(x,\dot{x},t)$ will, in general, change with time. The path taken by a mechanical particle must satisfy the principle of least action \cite{Cline},  which requires the first-order variation of the time integral of the Lagrangian to vanish for fixed end points. Yielding the Euler–Lagrange equation of motion. For a particle moving in one dimension, this equation takes the form
\begin{eqnarray} \label{EL}
    \frac{{\rm d}}{{\rm d}t}\left(\frac{\partial \mathcal{L}}{\partial \dot{x}}\right) - \frac{\partial \mathcal{L}}{\partial x} = 0 \, .
\end{eqnarray}
Despite the stark formal differences between Newtonian and Lagrangian mechanics, the two formalisms can be derived from one another \cite{Goldstein, gregory}. In fact, it is possible to reduce Eq.~(\ref{EL}) to Newton's second law in Eq.~(\ref{E1}) and vice versa. Lagrangian mechanics therefore does not introduce new physics, but rather provides a more convenient and coordinate-independent language for describing the same physical system. 

Nonetheless, the Lagrangian reformulation of CM brings with it a number of ontological shifts \cite{Butterfield} which, we argue, makes it less suitable for the physically motivated quantisation procedure. By renouncing an ontology of particles with precise position and velocity that interact causally via local forces, the theory provides a description in terms of less conceptually fundamental physical concepts, rendering it overall less perspicuous than Newtonian mechanics. For example, the idea that dynamical predictions for a mechanical particle can be made by considering its precise position and velocity at any given time $t$ is replaced by the abstract notion of a scalar function, namely the Lagrangian ${\mathcal L}$, which contains all the information about the mechanical system and its entire time evolution (cf.~e.g.~Eq.~\eqref{EL}). In Lagrangian mechanics, the ontological role played by particles in Newtonian mechanics is replaced by the possible paths taken by them. Moreover, since these paths must satisfy the least action principle, one might argue that the framework appears somewhat teleological, i.e.~it is more outcome-directed than Newtonian mechanics \cite{Glick}. 

\subsection{Hamiltonian mechanics}\label{sec:2.3}

Following the success of Lagrange's generalised description of classical mechanics, Hamilton \cite{Hamilton} tried to obtain an even more elegant mathematical structure by drawing inspiration from classical optics. When modelling the dynamics of mechanical particles moving in one dimension, this formalism replaces the velocity $\dot{x}$ with the conjugate momentum $p$ with
\begin{eqnarray}\label{ppp}
p = \frac{\partial \mathcal{L}}{\partial \dot{x}} \, ,
\end{eqnarray}
while retaining the position $x$ as the generalised coordinate. In Hamiltonian mechanics, the Hamiltonian $H$, rather than the Lagrangian $\mathcal L$, plays the central role. The Hamiltonian $H$ is constructed such that 
\begin{eqnarray} \label{Lana}
H=p\dot{x}-\mathcal{L}(x,\dot{x},t) \,.
\end{eqnarray}
This equation connects both quantities, $H$ and $\mathcal{L}$, through a Legendre transformation \cite{Goldstein,gregory}.
Provided that the momentum–velocity relation in Eq.~(\ref{ppp}) is invertible, the velocity $v=\dot{x}$ can be expressed in terms of $p$, thereby yielding a Hamiltonian $H(x,p,t)$ that depends no longer on the velocity $v$. If the Hamiltonian has no explicit time dependence, it can be shown to equal the total energy of the mechanical system,
\begin{eqnarray}
H(x,p) = T(p)+V(x) 
\end{eqnarray}
where $T(p)$ and $V (x)$ denote, again, the kinetic and the potential energy of the mechanical particle. In the case of a single particle moving along the $x$-axis, the equations of motion in Hamiltonian mechanics are \cite{Goldstein,gregory} 
\begin{eqnarray}\label{final}
    \dot x = \frac{\partial H(x,p)}{\partial p}, ~~~~
    \dot p =-\frac{\partial H(x,p)}{\partial x } \, . 
\end{eqnarray}
Notice that these two equations can be equivalently expressed in a more compact form using Poisson brackets $\{f,g\}$, which are   coordinate-independent \cite{Goldstein,gregory}. The Poisson bracket of two differentiable functions $f(x,p,t)$ and $g(x,p,t)$ equals 
\begin{eqnarray} \label{PB}
    \{f,g \} = \frac{\partial f}{\partial x}\frac{\partial g}{\partial p} - \frac{\partial f}{\partial p}\frac{\partial g}{\partial x} \, .
\end{eqnarray}
Using this notation, the time derivative of differentiable observables $f(x,p,t)$ can be written as
\begin{eqnarray}\label{poi}
    \frac{{\rm d}f}{{\rm d}t} = \{f,H\}+\frac{\partial f}{\partial t} \, .
\end{eqnarray}
Hence Poisson brackets can be used to identify conserved quantities. For example, applying Eq.~(\ref{poi}) and setting $f = g = H$ gives $\{H, H\} = 0$ which shows that the Hamiltonian $H$, if time independent, represents a conserved quantity. Moreover, one can show that the Poisson bracket 
$ \{x,p\}$ equals 
\begin{eqnarray}
    \{x,p\} = 1 \, . 
\end{eqnarray}
This relation serves as the classical counterpart of the canonical commutation relation, which is the starting point of canonical quantisation schemes.

Although the Hamiltonian formulation is mathematically the most elegant, the Legendre transformation in Eq.~(\ref{Lana}) which defines the Hamiltonian is not always reversible \cite{assumptions,HamvsLag}. In general, this is only the case if the Lagrangian is non-degenerate.\footnote{There are numerous extensions of Hamiltonian mechanics which can deal with such systems \cite{Goldstein, gregory, Brown}. However, these methods increase mathematical complexity, thereby reducing physical perspicuity.} The result can be a loss of generality in the description of mechanical particles. In addition, since Hamiltonian mechanics is derived from the Lagrangian framework, Hamiltonian mechanics carries many of the features which reduce the physical perspicuity of Lagrangian mechanics. Consequently, Hamiltonian mechanics provides a less perspicuous description than Newtonian mechanics.
Moreover, let us emphasize that Eq.~(\ref{ppp}) establishes a connection between the momentum $p$ and the velocity $v = \dot x$ and vice versa. This is not a problem, since adding a definition does not alter the predictions of a theory. However, when SQM promotes the momentum $p$ to the generator for spatial translation, the original physical meaning of the momentum, in particular, its link to velocity, becomes ambiguous, since its definition has changed.

\section{Quantum mechanics in configuration space} \label{sec:3}

In 2016, Bennett {\em et al.} \cite{Bennett} introduced a physically motivated quantisation approach for light in momentum space as a gauge-independent alternative to canonical quantisation approaches. This method starts by identifying all possible classical states of the system undergoing quantisation. Since the classical states are pairwise distinguishable, they are then directly promoted to pairwise orthogonal quantum states. This physically motivated quantisation approach allowed for a direct quantisation of photons in position space \cite{Jake,Daniel,AMC}. More recently, Bukhari {\em et al.}~\cite{Bukhari} proceeded analogously to quantise mechanical particles. On the grounds that the description given by Newtonian mechanics is physically the most perspicuous of the classical theories, it  became the starting point for QMCS. In this section, we review the resulting formalism.

\subsection{The quantum configuration space} \label{secsomething}

As we have seen in the previous section, to predict the dynamics of classical mechanical particles using Newtonian mechanics, we need to know their initial position $x$ {\em and} their initial velocity $v$. To ensure that both position and velocity play a similar role in the corresponding quantum theory, Ref.~\cite{Bukhari} promoted the classically distinguishable $(x,v)$ states to pairwise orthogonal quantum states $|x,v \rangle$. Assuming the same measurement postulates as in SQM and demanding that these states are also distinguishable in the quantum formalism implies that
\begin{eqnarray} \label{delta1}
\langle x,v| x',v' \rangle = \delta(x-x') \, \delta(v-v')
\end{eqnarray}
for all $x$, $x'$, $v$ and $v'$. 
Hence the $|x,v \rangle$ states form an orthonormal basis in the Hilbert space ${\cal H}$ of quantum mechanical particles. 

In contrast to classical particles, quantum mechanical particles can be prepared in a superposition of states associated with different positions and velocities. Their state vectors $|\psi(t) \rangle$ 
are of the same general form as the state vector in Eq.~(\ref{E1.2}) with the corresponding wave function $\psi(x,v,t)$ now given by $\psi(x,v,t) = \langle x,v |\psi(t)\rangle$. Adopting the same measurement postulates as SQM, the probability density for finding the mechanical particle at time $t$ at position $x$, while moving with velocity $v$, equals $\rho(x,v,t) = | \psi(x,v,t) |^2$ for normalised state vectors $|\psi(t) \rangle$ with $\langle \psi(t)|\psi(t) \rangle = 1$. Notice also that the $|x,v \rangle$ states are not the quantum states of individual particles, but rather the basis states from which single-particle states are built.

\begin{figure}[t]
\centering
\includegraphics[width=0.7 \linewidth]{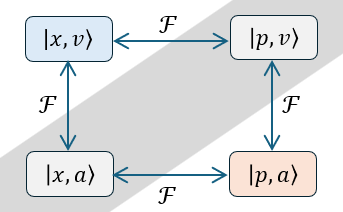}
\caption{Illustration of four different sets of basis states which are all linked via Fourier transforms and which can be used interchangeably to describe a single quantum mechanical particle moving along the $x$-axis. The most classical among the quantum basis states are the $|x,v \rangle$ states which correspond directly to the $(x,v)$ states of Newtonian mechanics. In contrast to this, the $|p,a \rangle$ states, are the least classical quantum bases, since they are equal superpositions of all $|x,v \rangle$ states. Such states do not exist in CM.} \label{fig:aa}
\end{figure}

Before looking at quantum mechanical observables, let us introduce three alternative representations of the state vector $|\psi(t) \rangle$ (cf.~Fig.~\ref{fig:aa}). An alternative sets of basis states are given by the $|p,v \rangle$, the $|x,a\rangle$ and the $|p,a\rangle$ states with
\begin{eqnarray} \label{E11}
&& |p,v\rangle = \frac{1}{\sqrt{2\pi \hbar}} \int_{\mathbb{R}}{\rm d}x \, {\rm e}^{{\rm i}px/\hbar} \, |x,v\rangle \, , \notag \\
&& |x,a\rangle = \frac{1}{\sqrt{2\pi \hbar}} \int_{\mathbb{R}}{\rm d}v \, {\rm e}^{{\rm i}av/\hbar} \, |x,v\rangle \, , \notag \\
&& |p,a \rangle = \frac{1}{2\pi \hbar} \iint_{\mathbb{R}^2}{\rm d}x \, {\rm d}v \, {\rm e}^{{\rm i}(px+av)/\hbar} \, |x,v\rangle \, .
\end{eqnarray}
Since these states relate to the $|x,v \rangle$ states via Fourier transforms, they are automatically pairwise orthogonal and form complete sets of basis states,
\begin{eqnarray} \label{xxkknn} 
&& \langle p,v|p',v'\rangle = \delta(p-p') \, \delta(v-v') \, , \notag \\
&& \langle x,a|x',a' \rangle = \delta(x-x') \, \delta(a-a') \, , \notag \\
&& \langle p,a|p',a' \rangle = \delta(p-p') \, \delta(a-a') \, .
\end{eqnarray}
For example, the $|p,a \rangle$ states describe particles that occupy all available positions $x$ and move at all available velocities $v$ at once. These states form a complementary basis to the $|x,v \rangle$ states. As we shall see below, these states are eigenstates of the momentum $\hat p$ and the acceleratum $\hat a$.  
\subsection{Basic observables and generators}

In QMCS, we construct observables in the same way as they are constructed in SQM \cite{Bukhari}. By definition, their eigenvalues are the possible measurement outcomes of the corresponding measurement. Moreover, their eigenstates are states which yield this result with maximum probability. Hence, $\hat x$ and $\hat v$ with
\begin{eqnarray}\label{xop}
&& \hat{x} = \iint_{{\mathbb{R}}^2} {\rm d}x\, {\rm d}v \, |x,v\rangle \, x \, \langle x,v| \, , \notag \\
&& \hat {v} = \iint_{{\mathbb{R}}^2} {\rm d}x\, {\rm d}v  \, |x,v\rangle \, v \, \langle x,v| 
\end{eqnarray}
are the position and the velocity observables of quantum mechanical particles. Analogously, using 
the definitions of the $|p,v \rangle$ and the $|x,a \rangle$ states in Eq.~(\ref{xxkknn}), we can introduce the two additional observables for the momentum $\hat p$ and the acceleratum $\hat a$,
\begin{eqnarray}\label{xop}
&& \hat{p} = \iint_{{\mathbb{R}}^2} {\rm d}p \, {\rm d}v \, |p,v\rangle \, p \, \langle p,v| \, , \notag \\
&& \hat {a} = \iint_{{\mathbb{R}}^2} {\rm d}x \, {\rm d}a  \, |x,a\rangle \, a \, \langle x,a| \, .
\end{eqnarray}
Exploring the properties of Fourier transforms in Eq.~(\ref{E11}), we find that the above multiplications with $p$ and $a$ correspond to derivatives in the $|x,v \rangle$ space such that \cite{Bukhari}
\begin{eqnarray} \label{E85b}
&& \hat p = -{\rm i}\hbar \iint_{{\mathbb{R}}^2} {\rm d}x \, {\rm d}v\,|x,v\rangle \, \frac{\partial}{\partial x} \, \langle x,v| \, , \notag \\
&& \hat a = -{\rm i}\hbar \iint_{{\mathbb{R}}^2} {\rm d}x \, {\rm d}v \, |x,v\rangle \, \frac{\partial}{\partial v} \, \langle x,v|
\end{eqnarray} 
with respect to the basis given by the $|x,v \rangle$ states. Moreover, Eq.~(\ref{E85b}) shows that $\hat p$ and $\hat a$ are not only observables, they are also generators. As in SQM, $\hat p$ is the generator of spatial translations. In addition, $\hat a$ generates changes in velocity. The reason for why we refer to $\hat a$ as the \emph{acceleratum} \cite{Bukhari} is the structural similarity between $\hat p$ and $\hat a$.  

To help intuitions as to what the above observables represent, we now take a closer look at the expectation values of the above observables for a given state $|\psi(t) \rangle$. For simplicity, we write the (complex) wave function $\psi(x,v,t)$ of this state in the following as
\begin{eqnarray} \label{E3.8extra}
\psi(x,v,t) = {\rm e}^{{\rm i} \phi (x,v,t)} \, \rho(x,v,t)^{1/2} 
\end{eqnarray}
with $\phi(x,v,t)$ representing phases.
Using this notation, we find that the momentum and the acceleratum expectation values $\langle p(t) \rangle = \langle \psi(t)|\hat p |\psi(t) \rangle$ and $\langle a(t) \rangle = \langle \psi(t)|\hat a |\psi(t) \rangle$ are given by 
\begin{eqnarray} \label{E3.11extra}
\langle p(t) \rangle 
&=& \hbar \iint_{{\mathbb{R}}^2} {\rm d}x\, {\rm d}v \left[ \rho(x,v,t) \, {\partial \over \partial x} \phi(x,v,t) 
\right. \notag \\
&& \left. - {{\rm i} \over 2} \, {\partial \over \partial x} \rho(x,v,t) \right] \, , \notag \\
\langle a(t) \rangle &=& \hbar \iint_{{\mathbb{R}}^2} {\rm d}x\, {\rm d}v \left[ \rho(x,v,t) \, {\partial \over \partial v} \phi(x,v,t) \right. \notag \\
&& \left. 
- {{\rm i} \over 2} \, {\partial \over \partial v} \rho(x,v,t) \right] 
\end{eqnarray}
with $\rho(x,v,t) = |\psi(x,v,t)|^2$. 
For example, mechanical particles with $\phi(x,v,t) = kx$ and constant $\rho(x,v,t)$ have the momentum expectation value $\langle p(t) \rangle = \hbar k$, as one would expect. Since the expressions in Eq.~(\ref{E3.11extra}) depend on the phase $\phi(x,v,t)$, they do not have a straightforward physical meaning in CM. This realisation fits well with the intense debates on how best to define momentum in classical physics (for more details see e.g.~Ref.~\cite{Goyal}). Nevertheless, in the case of particles which carry fields\textemdash such as photons, which are carriers of electric and magnetic field vectors\textemdash the momentum expectation value can be linked to field expectation values, as shown in Ref.~\cite{AMC}. These exist in classical electrodynamics, and can be measured. Considering particles with fields therefore offers an indirect way of measuring the momentum and the acceleratum of mechanical particles and gives both quantities a physical meaning. 

Given the many formal similarities, it is not surprising that SQM and QMCS yield the same commutator relation for the position and momentum operator. Using the above definitions, one can show that
\begin{eqnarray}\label{xpv}
     [\hat x, \hat p] = [\hat v, \hat a] = {\rm i} \hbar \, .
\end{eqnarray}
The above equation implies two uncertainty relations,
\begin{eqnarray} \label{S1zzz}
\Delta x \cdot \Delta p \geq \frac{\hbar}{2} \, , ~~~~
\Delta v \cdot \Delta a 
\geq \frac{\hbar}{2} \, .
\end{eqnarray}
Alongside the Heisenberg position–momentum uncertainty relation \cite{Sakurai}, we obtain a velocity–acceleratum uncertainty relation. Thus, in addition to the standard canonical pair $(\hat x,\hat p)$, quantum particles possess a second canonical pair $(\hat v,\hat a)$, with all other commutators vanishing,
\begin{eqnarray}\label{xvp}
     [\hat x, \hat v] = [\hat p, \hat a] = [\hat x, \hat a] = [\hat p, \hat v] = 0 \, . 
\end{eqnarray}
Hence, QMCS allows, for example, for simultaneous measurements of position and velocity. 

\subsection{Dynamics}

As shown in Ref.~\cite{Bukhari}, QMCS describes the dynamics of quantum mechanical particles with the help of a Schr\"odinger-type equation of the form 
\begin{eqnarray} \label{newschh}
{\rm{i}}\hbar \,\frac{\partial }{\partial t} |\psi(t) \rangle = \hat H_{\rm dyn} \, |\psi(t) \rangle 
\end{eqnarray}
with $\hat H_{\rm dyn}$ representing the dynamical Hamiltonian\footnote{Here we refer to this operator as a Hamiltonian, since it plays the same role as the Hamiltonian in SQM. However, as we shall see in Section \ref{sec4}, it is not related to the expression for the Hamiltonian in Hamiltonian mechanics, which is why we added the word "dynamical."} of the quantum mechanical particle. A closer look at the above equation shows that $\hat H_{\rm dyn}$ is the generator for time translation. Due to their close link to the $(x,v)$ states of Newtonian mechanics, the $|x,v \rangle$ basis states must evolve along well-defined classical trajectories in the quantum configuration space. Since the $|x, v\rangle$ states form a complete basis in the Hilbert space of the mechanical particles, this observation can be used to derive $\hat H_{\rm dyn}$ for any given potential $V(x)$ \cite{Bukhari}. Once the dynamical Hamiltonian is know, Eq.~(\ref{newschh}) can be used to predict the dynamics of any given initial quantum state. In the following Section, we illustrate this by studying a concrete example.

\section{A quantum mechanical particle in free space} \label{sec4}

To further illustrate the main features of QMCS and to further explore its relations with  alternative descriptions of mechanical particles, we now use this formalism to study the dynamics of quantum mechanical particles moving in one dimension in the absence of a potential $V(x)$. Let us begin by deriving the dynamical Hamiltonian $\hat H_{\rm dyn}$ in Eq.~(\ref{newschh}) for the case where $V(x)=0$. In free space, Newtonian mechanics tells us that the $|x,v \rangle$ states evolve such that\footnote{Notice that equation defines the phase factor of the $|x,v \rangle$ states, by assuming that these states evolve without changing their state compared to a given initial state, thereby removing all gauge-dependence from the formalism.} 
\begin{eqnarray} \label{However2}
   |x,v \rangle \longrightarrow |x+vt,v \rangle 
\end{eqnarray}
within a time interval $(0,t)$. Given $\psi(x,v,0)$ is the wave function of the initial state $|\psi(0) \rangle$, the quantum state $|\psi(t)\rangle$ at a later time $t$ therefore equals
\begin{eqnarray}\label{psittt00}
|\psi(t)\rangle &=&  \iint_{{\mathbb{R}}^2} {\rm d}x\, {\rm d}v  \, {\psi(x,v,0)} \, |x+vt,v \rangle \notag \\
&=&  \iint_{{\mathbb{R}}^2} {\rm d}x \, {\rm d}v  \, {\psi(x-vt,v,0)} \, |x,v \rangle \, .
\end{eqnarray}
Hence $\psi(x,v,t) = \psi(x-vt,v,0)$ which implies that
\begin{eqnarray}\label{psittt0000}
{\partial \over \partial t} \psi(x,v,t) = - v \; {\partial \over \partial x} \psi(x,v,t) \, .
\end{eqnarray}
This equation is closely linked to Eq.~(\ref{newschh}) and allows us to identify the corresponding generator for time translation. Given the definition of the momentum operator $\hat p$ in Eq.~(\ref{E85b}), we find that the dynamical Hamiltonian $\hat H_{\rm dyn}$ of a quantum mechanical particle moving in free space equals
\begin{eqnarray} \label{However}
   \hat H_{\rm dyn} = \hat{v} \hat{p} \, .
\end{eqnarray}
Since $\hat v$ and $\hat p$ commute, the above Hamiltonian is Hermitian and conserves the normalisation of state vectors evolving according to the Schr\"odinger equation. Indeed, one can check that the above Hamiltonian evolves the $|x,v \rangle$ states, as described by Eq.~(\ref{However2}), and quantum superposition states, therefore as described by Eq.~(\ref{psittt00}). 

Now we can use Heisenberg's equation of motion to analyse the dynamics of the expectation values $\langle A(t) \rangle = \langle \psi(t) |\hat A| \psi(t) \rangle
$ of any observable $\hat A$. For time-independent observables, Eq.~(\ref{newschh}) implies that 
\begin{eqnarray} \label{newschh222}
{\rm{i}}\hbar \, \frac{\rm d}{{\rm d} t} \langle A(t) \rangle = \big \langle \big[ \hat A, \hat H_{\rm dyn} \big] \big \rangle \, .
\end{eqnarray}
Using Eq.~(\ref{However}), we find, for example, that the position and velocity expectation values, $\langle x(t) \rangle$ and $\langle  v(t)\rangle$, evolve such that 
\begin{eqnarray}\label{xvEvolution}
\frac{\rm d}{{\rm d} t} \langle x(t) \rangle = \langle v(t)\rangle \, , ~~~~
\frac{\rm d}{{\rm d} t} \langle  v(t)\rangle =0 \, .
\end{eqnarray}
These two equations form a {\em closed} set of differential equations. This is not surprising, since the position-velocity space constitutes the configuration space of CM. Moreover, the time derivatives of the expectation values of momentum and acceleratum, $\langle p(t) \rangle$ and $\langle a(t) \rangle$, are given by
\begin{eqnarray}\label{paEvolution}
\frac{\rm d}{{\rm d} t} \langle p(t) \rangle =0  \, , ~~~~
\frac{\rm d}{{\rm d} t} \langle a(t) \rangle = - \langle p(t) \rangle \, .
\end{eqnarray}
These two equations form another {\em closed} set of differential equations. In QMCS, momentum and acceleratum provide a {\em complementary} description of mechanical particles, since the $|p,a \rangle$ states form an alternative basis to the $|x,v \rangle$ basis states. 

\subsection{The relation to Newtonian mechanics}

When comparing Eqs.~(\ref{E2}) and (\ref{xvEvolution}), we see that, in QMCS, the position and velocity expectation values, $\langle x(t) \rangle $ and $\langle v(t) \rangle $ of  mechanical particles in free space, evolve according to Newton's equation of motion, since $f(x)=0$ in this case.  
Eq.~(\ref{However2}) moreover shows that the $|x,v \rangle$ states of QMCS remain localised at all times. In this way, QMCS avoids the infinitely fast spreading of wave functions  associated with SQM \cite{Heger,Heger2}. The above observations illustrate that QMCS contains Newtonian mechanics as a limiting case (cf.~Fig.~\ref{fig:0}). The classical $(x,v)$ states of Newtonian mechanics can now be represented by the quantum states $|\psi \rangle$, like the ones in Eq.~(\ref{E1.2}), but with vanishing uncertainties in position $x$ and velocity $v$. Within QMCS, such states exist, because of the vanishing commutator between $\hat x$ and $\hat v$ in Eq.~(\ref{xvp}). This is not surprising, since the formalism of QMCS has been constructed accordingly \cite{Bukhari}. In other words, QMCS is indeed a reducing theory in respect to Newtonian mechanics.

Eqs.~(\ref{xvEvolution}) and (\ref{paEvolution}) moreover show that the velocity and the momentum expectation values of mechanical particles moving in free space are conserved, as one would expect in the presence of spatial and time translational symmetries \cite{Noether}. Furthermore, invariance under time translation implies conservation of the expectation value $\langle H_{\rm dyn}(t) \rangle$. However, this conservation remains hidden in CM, since the operator $\hat H_{\rm dyn} = \hat p \hat v$ does not have a corresponding classical observable, unless we consider particles which carry fields \cite{AMC}. Instead, CM notices that the energy of mechanical particles, which corresponds to the the quantum mechanical observable
\begin{eqnarray}\label{Hcm}
\hat H_{\rm CM} = \hat v^2/2m \, ,
\end{eqnarray}
is a conserved quantity. The expectation values of $\hat H_{\rm CM}$ are conserved also in QMCS, since $[\hat H_{\rm CM},\hat H_{\rm dyn}]=0$.

\subsection{The relation to standard quantum mechanics}

\renewcommand{\arraystretch}{1.2}
\begin{table*}[t]
\centering
  \begin{tabular}{|c|c||c|c|c|}
    \hline 
    & & QMCS & SQM & ~~~~ KvN ~~~~ \\
    \hline \hline
    ~~~ Observables ~~~ & Position & $\hat x$ & $\hat x$ & $x$ \\
     & ~Velocity ~ &  $\hat v$ & $-$ & $v$ \\
    \hline \hline
    Generators & Time translation ~ & ~~~ $\hat H_{\rm dyn}\sim \partial_t$ ~~~ & ~~~ $ \hat H_{\rm SQM} \sim \partial_t$ ~~~ & ~~~ $\lambda_t \sim \partial_t $ ~~~ \\
     & Position translation & $\hat p\sim \partial_x$ & $\hat p\sim \partial_x$ & $\lambda_x \sim \partial_x $ \\  
    & ~~~ Velocity translation ~~~ & $\hat a \sim \partial_v$ & $-$ & $\lambda_v \sim \partial_v $ \\
    \hline
  \end{tabular}
  \caption{A key difference between QMCS and SQM is that the former supports an additional observable and an additional generator, namely the velocity observable $\hat v$ and the acceleratum $\hat a$. QMCS therefore has more similarities with the classical formalism of KvN than with SQM, since KvN supports the same observables and the same generators. Nonetheless QMCS is a quantum and not a classical theory.}
  \label{table2}
\end{table*}

As we emphasize in Table \ref{table2}, the main difference between QMCS and SQM is the absence of a velocity observable $\hat v$ (which is independent of the momentum $\hat p$) and a generator for velocity changes, i.e.~the acceleratum $\hat a$, in the latter  formalism \cite{Griffiths,Sakurai,Liboff}. The reason for this is that SQM is based on Hamiltonian mechanics, which 
prioritises the conjugate momentum $p$ in Eq.~(\ref{ppp}) over the velocity $v$ of classical mechanical particles. More concretely, it is assumed that $p$ is a function of the velocity $v$ (cf.~Eq.~(\ref{ppp})) and momentum and velocity are therefore considered interchangeable variables. When SQM promotes the Poisson bracket in Eq.~(\ref{poi}) to the canonical commutation relation for $\hat x$  and $\hat p$, the momentum operator $\hat p$ becomes the generator for spatial translations in Eq.~\eqref{E85b}. As a result, it is no longer linked to velocity and SQM becomes unable to differentiate  between velocity and momentum. In contrast to this, in QMCS, velocity and momentum are two independent variables and $\hat v$ and $\hat p$ are commuting operators (cf.~Eq.~(\ref{xvp})). This applies, since we do not define the momentum $p$ with the help of Eq.~(\ref{ppp}); instead it represents the generator for spatial translations, is conserved in the presence of spatial translational symmetries \cite{Noether} and has no straightforward analogue in CM. 

Nevertheless, in situations where Ehrenfest's theorem holds \cite{Ehrenfest}, QMCS can be used to reproduce the predictions of SQM. One such example is a particle moving freely in one dimension. More concretely, to represent the  quantum states that are usually considered in SQM, QMCS needs to consider states $|\psi \rangle$ of the general form \cite{Davies}
\begin{eqnarray}
|\psi \rangle = \iint_{{\mathbb{R}}^2} {\rm d} x \, {\rm d} v \, \psi(x,t) \, |x,v \rangle \, .
\end{eqnarray}
Their wave functions $\psi(x,t)$ do not depend on $v$ and they are instead maximally de-localised in velocity \cite{Heger,Heger2}. Fortunately, the velocity degree-of-freedom is 
less relevant for the description of stationary particles, for which SQM is most commonly used, e.g. in the modelling of an electron inside a hydrogen atom. In contrast to this, QMCS delivers observables $\hat x$ and $\hat v$ for both position and velocity, as well as generators $\hat p$ and $\hat a$ for changes in those variables. To re-iterate the need for two independent observables $\hat v$ and $\hat p$ with $[\hat v,\hat p]=0$, let us take a closer look at photons, which are a special example of quantum mechanical particles \cite{Bukhari}. What distinguishes photons from massive particles is that they only move at one speed, namely the speed of light $c$ \cite{Jake,Daniel}. This means that the velocity operator $\hat v$ of photons only has the two eigenvalues $v = \pm c$. In contrast to this, the momentum of photons can assume any real value between $- \infty$ and $+\infty$. Since it is not linked to velocity, it is referred to as the \textit{moving} or the {\em dynamical} momentum \cite{AMC}. By making a clear distinction between the velocity $\hat v$ and the momentum $\hat p$, QMCS enhances the way in which wave-particle duality is implemented in quantum physics \cite{Bukhari}.

Since SQM does not support a velocity observable, its Hamiltonian $\hat{H}_{\rm SQM}$ for a quantum mechanical particle moving in free space cannot coincide with the dynamical 
Hamiltonian $\hat{H}_{\rm dyn}$ in Eq.~(\ref{However}). Coincidentally, the Schr\"odinger equation associated with the Hamiltonian 
\begin{eqnarray}\label{However3}
\hat{H}_{\rm SQM} = \hat{p}^2/2m \, ,
\end{eqnarray}
evolves position and momentum expectation values $\langle x(t) \rangle$ and $\langle p(t) \rangle$ in agreement with Hamilton's equation of motion in Eq.~(\ref{final}). These coincide with Newton's equations of motion when $V(x)=0$, if we assume that $p=mv$. However, as we have pointed out in the Introduction, it is not always possible to find a Hamiltonian $\hat{H}_{\rm SQM}$ which guarantees consistency with CM. Moreover, note that, in QMCS, particles which are initially localised in  position and velocity remain localised at all times, since their Hamiltonian is not bound from below and has positive as well as negative eigenvalues. This is not warranted in SQM,  rendering the quantum-classical transition of SQM conceptually problematic \cite{Heger,Heger2}, even in free space.

\subsection{The relation to Koopman-von Neumann mechanics}

Like QMCS, KvN \cite{Koopman,vonNeumann,Barandes} assumes that the configuration space of mechanical particles is their position-velocity space (cf.~Table \ref{table2}). Moreover, KvN  uses the state vectors $|\psi \rangle$ in Eq.~(\ref{E1.2}) to describe the states of mechanical particles \cite{Sudarshan,Mauro}. 
Using this notation, the expectation values $\langle A \rangle$ of observables $\hat A = \hat A(x,v)$, which only depend on $x$ and $v$, are given by $\langle \psi |\hat A|\psi \rangle$, as in QMCS. For the state vector in Eq.~(\ref{E1.2}), this expression yields
\begin{eqnarray}
\langle A \rangle 
= \iint {\rm d}x \, {\rm d}v \, A(x,v) \, |\psi(x,v,t)|^2 \, .
\end{eqnarray}
As in QMCS, KvN introduces generators for time, position and velocity translations (cf.~Table \ref{table2}). While the position and the velocity observables, $\hat x$ and $\hat v$, commute with each other (as in Eq.~(\ref{xvp})), some commutators of the generators with $\hat x$ and $\hat v$ are non-zero and relations analogous to Eq.~(\ref{xpv}) apply. It is therefore not surprising that the equations of motions in KvN and QMCS are essentially the same \cite{Sudarshan,Mauro}. 

However, KvN was never intended to be more than a {\em classical} theory. The interpretation of the state vectors $|\psi \rangle$ is therefore different from their interpretation in QMCS. While QMCS supports genuine quantum superposition states, $|\psi \rangle$ always represents a classical stochastic ensemble of mechanical particles in KvN. In both formalisms, $|\psi(x,v,t)|^2$ represents the probability density of finding one of those particles in the state $(x,v)$. Nevertheless, representing a classical statistical mixture, the expectation value $\langle A \rangle $ of a state vector  $|\psi \rangle = c_1 \, |\psi_1 \rangle + c_2 \, |\psi_2 \rangle$ in KvN must be such that 
\begin{eqnarray} \label{oma}
\langle A \rangle
= |c_1|^2 \, \langle A \rangle_1 + |c_2|^2 \, \langle A \rangle_2 
\end{eqnarray}
with $\langle A \rangle_i = \langle \psi_i| \hat A |\psi_i \rangle$. In general, this is not the case. Using the relation
 $\langle A \rangle = \langle \psi |\hat A|\psi \rangle$, we find that 
\begin{eqnarray} \label{opa}
\langle A \rangle
= |c_1|^2 \, \langle A \rangle_1 + |c_2|^2 \, \langle A \rangle_2 + c_1c_2^* \, \langle \psi_1| \hat A |\psi_2 \rangle + {\rm c.c.} \notag \\
\end{eqnarray}
which is not the same as Eq.~(\ref{oma}). To enforce its classical nature, KvN therefore relies on a super-selection rule \cite{Sudarshan,Mauro} which only supports states which obey the relation in Eq.~(\ref{oma}) and prevents us from modelling genuine quantum superposition states. In the language of modern quantum physics, statistical mixtures of classical $(x,v)$ states are represented by the density matrices $\rho$ of the specific form in Eq.~(\ref{E1.3}). Considering such density matrices, QMCS can reproduce the predictions of KvN, as illustrated in Fig.~\ref{fig:0}. However, unlike KvN, it also supports the density matrices $\rho$ of the much more general form in Eq.~(\ref{E1.4}), which can exhibit quantum interference effects, like the ones illustrated in Eq.~(\ref{opa}), in case of measurements, \cite{Davies}.

\section{Conclusions} \label{sec5}

In general, the coexistence of multiple theoretical formulations describing the same mechanical system enhances our understanding because each one offers additional mathematical structure and complementary insights. Different theoretical formulations have different levels of complexity, different ranges of validity and therefore also different applications. One response to such variety is to take one specific quantum theory as fundamental, and then try to identify a particular formulation of CM as a preferred emergent description. A different response, the one which we have advocated here, is to motivate the choice of a particular quantum theory in terms of the classical theory which it quantises. The characteristic quantities, like position and velocity, of the classical theory will then have direct analogues in the quantum theory.

This paper illustrates that starting from a more physically perspicuous formulation of CM can result in a more perspicuous formulation of quantum theory. Since we consider Newtonian mechanics the most perspicuous, most conceptually transparent formulation of CM\textemdash the one whose terms have the most direct physical interpretation\textemdash we have endorsed the quantum theory which results from quantising the characteristic quantities of Newtonian mechanics. We have called this style of argument “physically motivated quantisation”. Physically motivated quantisation \cite{Bennett,Jake,Daniel} is a specific instance of the more general methodological approach which we laid out in the introduction: having our choice of more-ontologically-fundamental theory heuristically driven by our preferred less-ontologically-fundamental theory. This approach therefore increases the conceptual continuity between a given formulation of CM and its reducing quantum theory \cite{Bukhari}.

There is a clear sense in which the approach that we have taken in this paper involves identifying the quantities and concepts of classical physics as being less ontologically fundamental than those of quantum physics, yet as being more conceptually fundamental than their quantum counterparts. Classical quantities and concepts are not ontologically fundamental, in that we understand CM as being reducible to QM. However, the quantities and concepts of CM in its Newtonian formulation remain conceptually fundamental. These quantities and concepts are acting as a fixed point in our route to the underlying quantum theory: the adequacy of the quantum theory is assessed based on the extent to which it can successfully underwrite the original classical quantities and concepts. The physically motivated quantisation approach, therefore, counts continuity with prior physical concepts as an essential part of the justification of a proposed more fundamental framework\textemdash and, the more direct this continuity, the better.

The formulation of SQM is widely seen as one of the greatest triumphs of modern physics \cite{Griffiths,Sakurai,Liboff}. However, many highly successful theories in quantum physics, like quantum chemistry, quantum optics, and condensed matter physics, do not closely resemble SQM. Instead, they are often based on phenomenological models \cite{Hubbard}, rely on the presence of environment-induced decoherence mechanisms \cite{Zurek,Brody,Brukner} or apply a wide range of standard approximations. For example, quantum chemistry introduces so-called Ehrenfest trajectories in the numerical simulations of complex molecules \cite{Shalashilin}. Moreover, condensed matter physics often employs Hamiltonians which are not bound from below. Mathematically speaking, these Hamiltonians have more in common with the dynamical Hamiltonians of QMCS than with the positive Hamiltonians of SQM. 

A potential reason why physically motivated quantisation approaches have not been considered previously is that many traditional quantum experiments, like spectroscopy experiments on atoms and molecules, are performed on quantum systems prepared in stationary states. Indeed, many quantum physics textbooks only consider the energy eigenstates of quantum mechanical particles in trapping potentials, ranging from square well potentials and the harmonic oscillator to the electron in a hydrogen atom \cite{Sakurai,Griffiths}. In these cases, we are only interested in the stationary solutions of the time-independent Schr\"odinger equation, whose position expectation value remains constant in time. The velocity degree of freedom does not contribute to the modelling of such systems and is therefore easily overlooked.

When a physically-motivated quantisation procedure is applied to mechanical particles, it leads directly to QMCS. A central observation of QMCS is that position and velocity are placed on the same footing as independent variables, each carrying its own conjugate variable and canonical structure, made precise by the two commutation relations in Eq.~(\ref{xpv}). Moreover, the free evolution in QMCS reveals a richer structure than that of SQM \cite{Davies}. Consequently, QMCS might simplify the modelling of physical systems which evade straightforward quantum  descriptions, like relativity, gravity, chemistry and chaos, since it might allow us to approach these topics in a more direct way.   \\[0.5cm]
{\em Acknowledgement.}  AB thanks the Saudi government ministry of education represented by the Saudi Arabian Cultural Bureau in London (SACB) for their financial support. Moreover, MJD thanks Jonny Downes for helpful discussions.

\end{document}